\documentclass[a4paper,11pt,fleqn]{article}\usepackage[]{graphicx}\usepackage[table, usenames, dvipsnames]{xcolor}
\makeatletter
\def\maxwidth{ %
  \ifdim\Gin@nat@width>\linewidth
    \linewidth
  \else
    \Gin@nat@width
  \fi
}
\makeatother

\definecolor{fgcolor}{rgb}{0.345, 0.345, 0.345}

\usepackage{framed}
\makeatletter
 {\par\unskip\endMakeFramed%
 \at@end@of@kframe}
\makeatother

\definecolor{shadecolor}{rgb}{.97, .97, .97}
\definecolor{messagecolor}{rgb}{0, 0, 0}
\definecolor{warningcolor}{rgb}{1, 0, 1}
\definecolor{errorcolor}{rgb}{1, 0, 0}
\newenvironment{knitrout}{}{} 

\usepackage{alltt}
\usepackage{amsmath}
\usepackage[T1]{fontenc}
\usepackage[utf8]{inputenc}
\usepackage{cmbright}
\usepackage[margin=15pt,font=small,labelfont=bf]{caption}
\usepackage{booktabs}
\usepackage[margin=25mm]{geometry}
\usepackage{setspace}
\usepackage[dvipsnames]{xcolor}
\definecolor{mylinkcolour}{HTML}{FC8D62}
\definecolor{myurlcolour}{HTML}{8DA0CB}
\definecolor{mycitecolour}{HTML}{66C2A5}
\usepackage{textcomp}
\usepackage{amssymb}
\usepackage{capt-of}
\usepackage{bm}
\usepackage{accents}
\usepackage{stfloats}
\usepackage[backend=biber,style=authoryear,natbib=true,url=false,doi=true]{biblatex}
\bibliography{refs,packages}
\usepackage{xr}
\usepackage{dsfont}
\usepackage[section]{placeins}
\usepackage{longtable}

\usepackage{amsthm}
\theoremstyle{definition}

\usepackage{lineno}
\usepackage{graphicx}
\DeclareGraphicsExtensions{.pdf,.PDF,.png,.PNG}

\usepackage{authblk}
\author[1]{Stephen E Lane}
\author[2]{Richard Gao}
\author[2]{Matthew Chisholm}
\author[1]{Andrew P Robinson}
\affil[1]{Centre of Excellence for Biosecurity Risk Analysis, University of Melbourne, Parkville, Victoria 3010, Australia, \texttt{lane.s@unimelb.edu.au}}
\affil[2]{Department of Agriculture and Water Resources, Canberra, Australian Capital Territory 2601, Australia}
\usepackage{hyperref}
\hypersetup{
  linkcolor = mylinkcolour,
  urlcolor = myurlcolour,
  citecolor = mycitecolour,
  colorlinks = true,
  pdfauthor = {Stephen E Lane},
  pdftitle = {Statistical profiling to predict the biosecurity risk presented by non-compliant international passengers}
}

\providecommand{\tightlist}{%
  \setlength{\itemsep}{0pt}\setlength{\parskip}{0pt}}
\providecommand{\keywords}[1]{\textbf{\small Keywords:} \small #1}

\date{\today}
\title{Statistical profiling to predict the biosecurity risk presented by non-compliant international passengers}
\IfFileExists{upquote.sty}{\usepackage{upquote}}{}
\begin{document}

\maketitle

\clearpage

\begin{abstract}
Biosecurity risk material (BRM) presents a clear and significant threat to national and international environmental and economic assets. Intercepting BRM carried by non-compliant international passengers is a key priority of border biosecurity services. Global travel rates are constantly increasing, which complicates this important responsibility, and necessitates judicious intervention. Selection of passengers for intervention is generally performed manually, and the quality of the selection depends on the experience and judgement of the officer making the selection. In this article we report on a case study to assess the predictive ability of statistical profiling methods that predict non-compliance with biosecurity regulations using data obtained from regulatory documents as inputs. We then evaluate the performance arising from using risk predictions to select higher risk passengers for screening. We find that both prediction performance and screening higher risk passengers from regulatory documents are superior to manual and random screening, and recommend that authorities further investigate statistical profiling for efficient intervention of biosecurity risk material on incoming passengers.
\end{abstract}

\keywords{Biosecurity, risk prediction, profiling, international air passengers}

\clearpage

\section{Introduction}
\label{introduction}

Increasing globalisation and international travel increase the risks of pests and diseases invading novel areas and systems \citep[e.g.,][]{Hulme2009}
. Passengers arriving at international borders represent biosecurity risk because they may carry pests and/or diseases that are transmitted by biosecurity risk material (BRM), for example fruit from the originating country may be infested by fruit flies, and uncooked meat articles may be carrying Foot and Mouth Disease (FMD).

The impact of invasive pests and diseases is economically and environmentally substantial. The annual cost of invasive weeds in Australia has been estimated as AUD\$4bn \citep{Sinden_2005-yv}, and of invasive species generally to over USD\$200bn \citep{Pimentel2011-pe} annually. Furthermore, invasions can lead to a loss of ecosystem diversity --- for example, invasive alien species were estimated to decrease local diversity by 51\% \citep{Vila2011-uh}. Invasives are cited by the IUCN as being the most significant threat to biodiversity in islands, and the second most significant after climate change elsewhere \citep{iucn-2016}.

The Department of Agriculture and Water Resources (the department) has primary responsibility for managing the biosecurity system in Australia \citep{Department_of_Agriculture_and_Water_Resources_undated-kv}. The department undertakes interventions of various kinds in the many pathways over which it has regulatory authority, based on the principle that prevention is better than cure \citep[e.g.,][]{Leung2002-dp}. For example, the department screens incoming passengers at clearance points (e.g., sea and airports) for BRM. This screening process typically involves assessment of the passenger's paperwork, an interview, and possibly examination of the passenger's effects by x-ray or a Detector Dog Unit, or physical inspection.  The screening process is focused on detecting undeclared BRM only; declared BRM can be intercepted without screening, because it is voluntarily presented for inspection.

Screening all international passengers with equal effort is inefficient because different cohorts of passengers present different risks. Profiling of some form is generally involved in screening passengers for BRM both in Australia and internationally. Profiling is generally performed manually, whereby the officer assesses passengers to decide on how they should be processed. The quality of this assessment depends on the experience and judgement of the officer.

Increasing numbers of passengers and greater diversity in originating countries passing through Australia's borders mean that the profiling task for officers is becoming more complex. For example, the number of international passenger clearances increased by 4\% from 2013--14 to 2014--15 and 6\% from 2014--15 to 2015--16 \citep{Department_of_Agriculture_and_Water_Resources2016-th}. If passengers can be classified into groups or cohorts that reflect varying degrees of risk of non-compliance, then focusing intervention efforts upon those with higher risk should result in a higher rate of detection of non-compliance, whilst at the same time expending the same or fewer resources \citep{Cannon2009-ua}.

Expert judgement for allocating passengers into groups of varying non-compliance risk can be time consuming and may be inefficient \citep{Burgman2016-jt}. An alternative approach is to augment the allocation process by using a statistical profiling tool. Given a cohort of arriving passengers and associated traits/demographics, along with a risk prediction rule that places these passengers into various risk groups, a selection strategy can be defined to maximise detection of non-compliance.

Research into risk-based sampling for border intervention has historically focused on goods and commodities entering through ports. For example, \citet{Robinson2011-aw} investigated risk-based allocation of inspection resources for unit load devices (air transportation containers), and \citet{Hua2005-pr} used a combination of logistic regression and clustering to define a rule-based risk decision system for inspection of goods entering China. We are not aware of any reported research into the performance of statistical profiling for risk-based selection of arriving passengers for the purpose of detecting BRM. \citet{De_Melo2014-gp} investigated associations of air passenger traits with possession of BRM, but did not look further at the possibilities of using this information for profiling. Related but more specific investigations were undertaken by \citet{Lin2009-ny}, who constructed a model to predict FMD status in meat illegally carried by air passengers, and \citet{Lai2012-cv} who constructed a similar model for avian influenza virus. \citet{Shih2005-bk} investigated associations of passenger traits with monthly non-compliance counts, but did not investigate risk prediction for individual passengers.

We have two aims for the present article: (i) to assess the predictive ability of statistical profiling methods for predicting non-compliance with biosecurity regulations using passenger traits recorded on regulated documentation such as Incoming Passenger Cards (IPCs); and (ii) to evaluate the performance of the risk predictions in terms of screening higher risk passengers. We performed assessments via a cross-validation study using inspection data, which we describe in Sections \ref{sec:data} (data) and \ref{sec:methods} (methods). We present the results of the study in Section \ref{sec:results}, and make closing remarks in Section \ref{sec:discussion}.

\section{Data}
\label{sec:data}

Staff at Kingsford-Smith International Airport in Sydney, Australia conducted a census operation on a particular arriving flight between 18 June and 20 August 2015 inclusive. The flight was chosen due to a relatively high incidence of BRM interceptions, and a desire to minimise unnecessary screening activity by improving profile specificity. We used the data collected during this operation to perform a detailed analysis of passenger non-compliance. During normal operations, many passengers are not subjected to inspection, and passenger details are only recorded when undeclared biosecurity risk material (BRM) is found.

During the census operation all passengers who arrived were screened. Screening in this context refers to passenger's luggage being examined, e.g., by X-ray and/or opening and physically examining the luggage contents. Passenger traits were recorded for all passengers, along with details of any BRM located; passengers found with undeclared BRM were labelled as being non-compliant. The dataset comprised 3361 records, of which 6.5\% were non-compliant. Table \ref{tab:Vars} shows the passenger traits that were recorded. Cross-classification tables cannot be published for privacy reasons.

\begin{table}[!htp]
  \centering
  \caption{Description of passenger traits recorded during the screening process. Stage 1 predictor variables are those that can be used for profiling without direct reference to the IPC, whilst Stage 2 predictor variables require the IPC. See Section~\ref{prediction-models-compared} for modelling details.}
  \label{tab:Vars}
  \begin{tabular}{p{0.3\linewidth}p{0.65\linewidth}}
    \toprule
    \textbf{Non-compliance status} & A binary variable, with value 1 for a non-compliant result and 0 for a compliant result in the screening process (where `non-compliant' indicates the presence of any undeclared BRM). \\
    \multicolumn{2}{l}{\textit{Stage 1 predictor variables}} \\
    \textbf{Year of birth/age} & A numerical variable recording the passenger's year of birth. For ease of interpretation, this was transformed to age at the time of screening (years). \\
    \textbf{Sex} & A binary variable, with values \textit{male} and \textit{female}. \\
    \multicolumn{2}{l}{\textit{Stage 2 predictor variables}} \\
    \textbf{Citizenship group} & A categorical variable indicating the passenger's citizenship, as found on the IPC, and grouped into geographical regions. \\
    \textbf{Declaration status} & A binary variable, with value 1 for a declarant and 0 for a non-declarant, based on the passenger's Incoming Passenger Card (IPC). \\
    \textbf{Occupation} & A categorical variable for the passenger's occupation, as recorded on the IPC. \\
    \textbf{Visit reason} & A categorical variable for the passenger's visit reason, as found on the IPC. \\
    \bottomrule
  \end{tabular}
\end{table}

\subsection{Data preprocessing}
\label{data-preprocessing-and-exploration}

Some processing of the dataset was required before model fitting. Some levels within passenger traits were sufficiently rarely observed that we decided to collapse these into a \emph{not otherwise specified} level; specifically, levels that contained less than 50 observations were collapsed together prior to formal statistical modelling.

\section{Methods}
\label{sec:methods}

Repeated tenfold cross-validation was used for model comparison in this
study, with ten repeats, resulting in 100 training/testing datasets for
comparison.

\subsection{Prediction models compared}
\label{prediction-models-compared}

For each training/testing dataset, the following models were fit. Brief
descriptions are provided below, with more detail in Appendix~\ref{sec:appendix-mods}.

\begin{enumerate}
\def\labelenumi{\arabic{enumi}.}
\tightlist
\item semi-parametric logistic regression (\textbf{GAM})
\item a random forest with the number of trees optimised (\textbf{RF-caret})
\item a gradient boosting machine (GBM) with custom feature selection/collapsing and tuning parameter selection (\textbf{GBM-custom})
\item a GBM with optimised tuning parameters (interaction depth, number of trees/base learners, learning rate; \textbf{GBM-caret})
\item a neural network with the number of units in the hidden layer, and the weight decay optimised (\textbf{NN-caret})
\item a Bayesian (shrinkage) logistic regression models with normal priors (\textbf{Bayes-normal})
\item a Bayesian lasso model with Laplace priors (\textbf{Bayes-lasso})
\end{enumerate}

The random forest \citep{Breiman2001-yt}, GBM \citep{Friedman2001-dh} and neural network \citep{Ripley2008-nx} models were chosen due to their popularity in machine learning, ease of implementation in \texttt{R}, and well-documented performance \citep[e.g.,][]{Hastie2009-ns}.

Random forests are a machine-learning technique that perform \textit{ensemble learning} for classification, where the ensemble consists of a large number of classification trees (known as base learners). Each base learner is trained on a random sample of the data, and is grown to the maximum possible extent. Growing the base learner to the maximum possible extent results in overfitting, where predictions on a new dataset perform poorly, hence the resampling and refitting of multiple base learners. To make a prediction for a new observation, each base learner classifies that observation. The classification that occurs the most (over all trees in the forest) is taken to be the prediction for the observation.

The GBM also provides ensemble learning, but the base learners in a GBM are \textit{weak} learners; they are not grown to the maximum possible extent. The GBM starts with an imperfect model for the data (i.e. the base learner that is not grown maximally), and constructs a new model by successively fitting the residuals of the current model, using the same class of base learners as the initial imperfect model.

Neural network (NN) models are inspired by biological networks within the brain, in which messages are passed between neurons to send instructions for various biological processes. In NN machine learning, the \textit{messages} are the weighted predictor variables, which are passed along the network until they reach the output variables. Due to the many paths a message can take through the network, NN models provide a very flexible way of non-linear model fitting.

Machine learning algorithms require custom programming to be implemented in most operational environments, so we decided to also compare logistic regression \citep[with a semiparametric model, e.g.][]{Wood2011-xu}, which can be more readily implemented.

The Bayesian shrinkage methods \citep{Park2008-la} were chosen as a compromise between the machine learning and logistic regression methods. With some passenger traits being rarely observed (Section \ref{sec:data}), we felt these models would provide robustness against any separation issues that may occur.

Further, for each method we fit models with two different passenger trait sets. In the first stage, we use passenger traits that could be identified manually by the officer, without reference to the IPC. Profiles formed from this grouping could be used quickly by officers in the facility to select passengers for further screening. Models in this stage use age and as predictors, shown under the sub-heading \textit{Stage 1 predictor variables} in Table~\ref{tab:Vars}. The second stage uses the first stage passenger traits, along with those listed under the sub-heading \textit{Stage 2 predictor variables} as given in Table~\ref{tab:Vars}. The extra passenger traits, citizenship group, declaration status, occupation and visit reason all require inspection of IPCs, and are thus more time consuming. The second stage models possibly provide better predictions however, so we compare both stages for each model.

\subsection{Prediction metrics}
\label{prediction-metrics}

\subsubsection{Overall model comparison}
\label{sec:over-model-comp}

Overall model predictive ability was compared using the area under the receiver operator curve (AUC; e.g. \cite{Hanley1982-qn}), and predictive log-loss (e.g. \cite{Winkler1967-qy}). AUC is the curve created by plotting the true positive rate against the false positive rate (in the testing dataset) at various threshold probabilities, as if every passenger in the testing dataset were screened for non-compliance. Larger AUC values are associated with models that have better predictive ability.

Predictive log-loss measures how well the probabilities of non-compliance are predicted\footnote{In contrast to AUC, which measures the effect of classification at various thresholds.}. Given a sample of $n$ passengers, let $y_{i} = 1$ if a passenger is non-compliant and 0 otherwise, and $\hat{p}_{i}$ be the predicted probability that a passenger is non-compliant. Then, predictive log-loss is defined as $-(1/n)\sum \left[y_{i} \log(\hat{p}_{i}) + (1 - y_{i})\log(1 - \hat{p}_{i})\right]$. The $\log$ penalises `confident' predictions (close to 0 or 1) that are incorrect. Smaller values of predictive log-loss are associated with models that produce better predictions.

\subsubsection{Passenger screening comparisons}
\label{sec:model-comp-with}

In practice, it is prohibitively expensive and slow to inspect every passenger, and so some form of screening selection is required. As described earlier, manual profiling is performed by officers who select passengers for screening.  We simulated a targeted screening strategy that uses the passenger's predicted probabilities of non-compliance to select those that are to be screened. We perform this screening step on the withheld testing datasets described above. In this targeted strategy, passengers are ordered according to predicted risk of non-compliance with the top $P_{\text{scr}}\%$ of passengers all screened. In practice, we would also randomly select a proportion of the remaining passengers to keep current information about apparently low-risk cohorts; we omit this sample in the current exercise.

Some commonly reported metrics for these classification tasks are the positive predictive value, which is the proportion of selected passengers who were truly non-compliant, and the true positive rate, which is the proportion of non-compliant passengers selected. It is however, informative to consider the results in terms of \textit{efficiency}, which is the value of, for example, positive predictive value at a fixed screening proportion, relative to what would be achieved under a random screening approach. This approach has the advantage that the various metrics give equivalent estimated efficiency, simplifying the comparison between methods. Further, we can compare these efficiency estimates to those estimated for manual profiling from previous data, as we now describe.

As part of performance measurement, the Department regularly conducts post-screening inspection surveys to estimate the efficiency of the manual profiling by officers\footnote{We are unable to show this data due to privacy reasons.}. Results from the most recent survey data show that the estimated non-compliance rate during the current census was less than that estimated by the survey. However, if we assume that the relative efficiency of manual profiling to the overall non-compliance rate (equivalent to random profiling, as discussed above) is the same in the current data collection as it was during the survey, then we can compare manual profiling performance to statistical profiling performance via efficiency. During the recently conducted survey, manual profiling by officers resulted in an estimated efficiency of 1.3 as compared with random sampling.

\section{Results}
\label{sec:results}

\subsection{Overall predictive ability}
\label{overall-predictive-ability}

Figure~\ref{fig:modsAUC} displays boxplots of the AUC and predictive log-loss for all methods. The performance of the \textbf{GBM-custom} and \textbf{RF-caret} were extremely poor in comparison to the other methods based on predictive log-loss, so we do not consider these methods any further. There is little to separate models fit using Stage 1 or Stage 1 and 2 passenger traits based on predictive log-loss. Comparisons with AUC however show that including all passenger traits in the models improves predictive ability. Similarly, there is little difference between methods, with \textbf{GBM-caret} having the highest (although not statistically significantly so) median AUC and lowest log-loss.

\begin{knitrout}
\definecolor{shadecolor}{rgb}{0.969, 0.969, 0.969}\color{fgcolor}\begin{figure}[!hbp]

{\centering \includegraphics[width=0.85\textwidth]{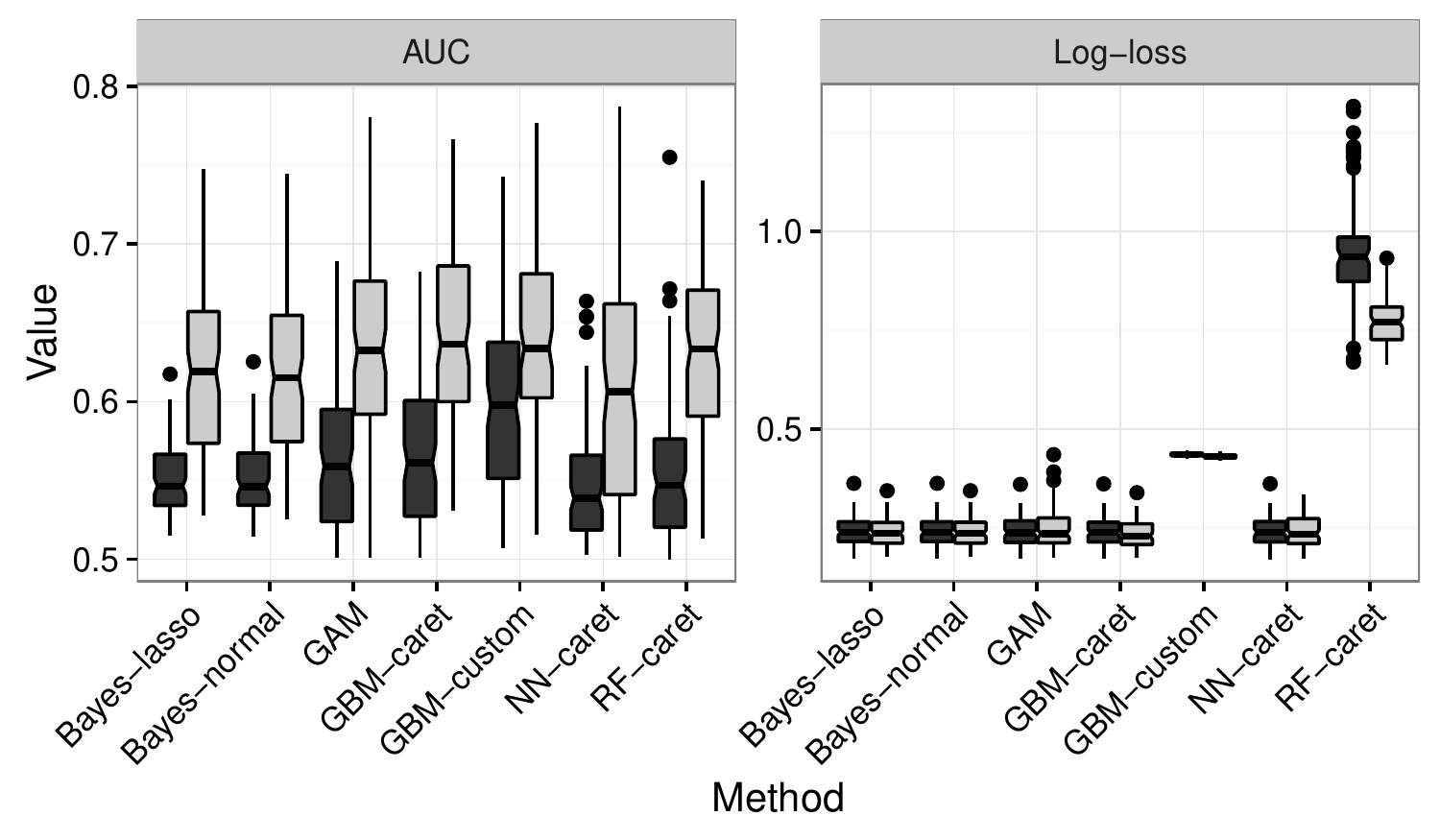} 

}

\caption[Comparison of AUC and predictive log-loss between methods]{Comparison of AUC and predictive log-loss between methods. Dark grey fill are from models using Stage 1 passenger traits, whilst light grey fill are from models using Stage 1 and 2 passenger traits.}\label{fig:modsAUC}
\end{figure}

\end{knitrout}

\subsection{Passenger screening performance}
\label{passenger-screening-performance-metrics}

Figure~\ref{fig:modsHitRateInd} shows efficiency as a function of screening rate. Panels in the figure represent the prediction method, and the solid black line is from models using Stage 1 passenger traits whilst the dashed black line is from models using Stage 1 and 2 passenger traits. Models fitted using Stage 1 and 2 passenger traits outperform those fitted using only Stage 1 passenger traits for all methods except \textbf{NN-caret} from very low screening rates, until approximately 60--70\% of passengers are screened.

Also shown in Figure~\ref{fig:modsHitRateInd} is the efficiency of manual profiling by officers (1.3), as discussed in Section~\ref{sec:model-comp-with}. It is clear from this figure that profiles created from statistical modelling (using Stage 1 and 2 passenger traits) outperform manual profiling by up to 50\% in efficiency.

\begin{knitrout}
\definecolor{shadecolor}{rgb}{0.969, 0.969, 0.969}\color{fgcolor}\begin{figure}[!hbp]

{\centering \includegraphics[width=0.85\textwidth]{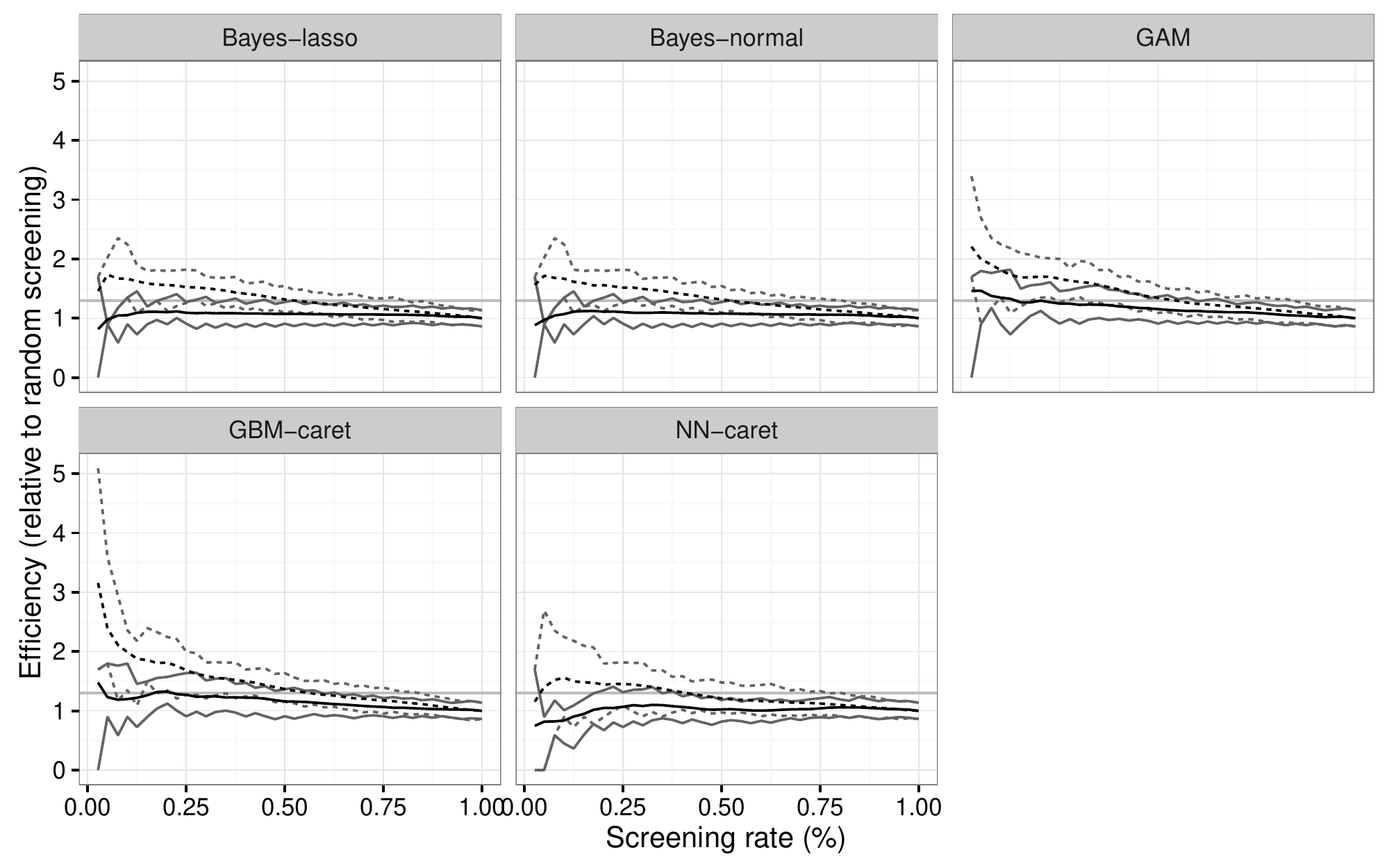} 

}

\caption[Comparison of efficiency as a function of screening rate for each method]{Comparison of efficiency as a function of screening rate for each method. The solid black line is from models using Stage 1 passenger traits, and the dashed black line is from models using Stage 1 and 2 passenger traits. 50\% pointwise intervals are shown in grey. The horizontal grey line shows the estimated efficiency of manual profiling.}\label{fig:modsHitRateInd}
\end{figure}

\end{knitrout}

\subsection{Importance of passenger traits}
\label{sec:import-pass-traits}

To assess the importance of individual passenger traits, we fit a model to the full dataset using both Stage 1 and 2 passenger traits. We chose to use the \textbf{GBM-caret} method due to its better performance as demonstrated\footnote{Whilst not significantly superior, it has the benefits of flexibility and ease of implementation as discussed in Section~\ref{prediction-models-compared}.} by Figures~\ref{fig:modsAUC} and \ref{fig:modsHitRateInd}.

Figure~\ref{fig:varImp} displays the relative importance of the passenger traits. Importance as calculated in the \texttt{GBM} package is the relative influence a passenger trait has on reducing the loss function used in the GBM. Importance was standardised so the sum of all estimated passenger trait influences equalled 100. Age, occupation and the reason for the visit are estimated as being much more important for predicting non-compliance than a passenger's declaration status, citizenship group or sex.

\begin{knitrout}
\definecolor{shadecolor}{rgb}{0.969, 0.969, 0.969}\color{fgcolor}\begin{figure}[!hbp]

{\centering \includegraphics[width=0.85\textwidth]{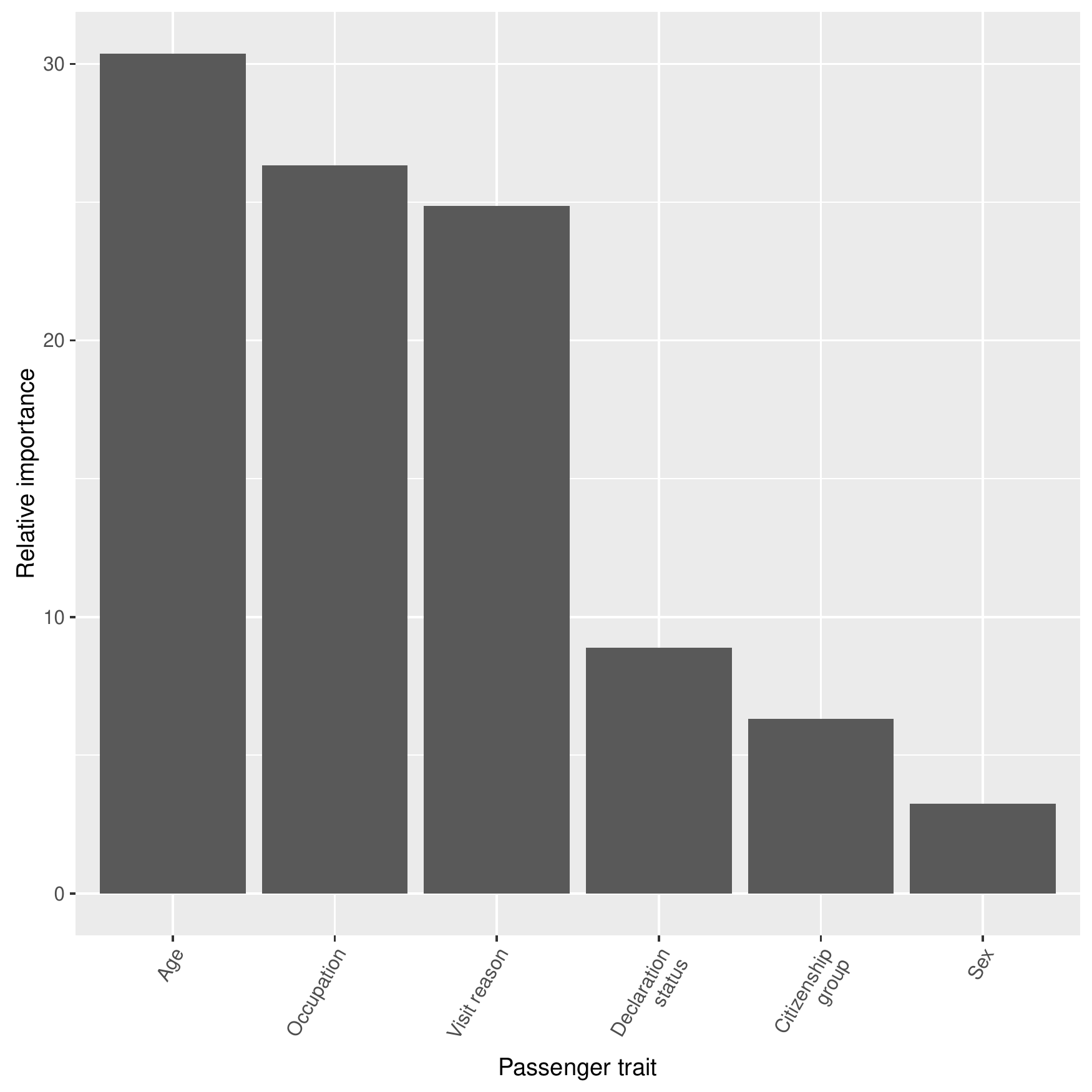} 

}

\caption[Relative importance of passenger traits in predicting biosecurity risk material non-compliance in a GBM model fit to the full dataset]{Relative importance of passenger traits in predicting biosecurity risk material non-compliance in a GBM model fit to the full dataset.}\label{fig:varImp}
\end{figure}

\end{knitrout}

Figure~\ref{fig:margEffect} displays the marginal effect of the three passenger traits\footnote{Levels of the traits have been masked due to privacy reasons.} with the largest relative importance (Figure~\ref{fig:varImp}); marginal effects being calculated using the tree traversal method \citep{Friedman2001-dh}. Age has a non-linear effect on non-compliance, with large relative jumps at around 20 and 45 years of age, and a large fall at around 60 years of age. Also shown in the occupation and visit reason panels are the number of passengers with each level of the trait, which provides a gauge of the precision related to the estimated marginal effect.

A demonstration of the ability of the GBM method to pick up interaction effects is shown in the bottom right panel of Figure~\ref{fig:margEffect}. This panel shows the interaction effect of occupation and citizenship grouping; citizenship groupings are represented by different line types. The effect of occupation in passengers who are citizens of country Y is clearly different to that of passengers from the other two countries of citizenship.

\begin{knitrout}
\definecolor{shadecolor}{rgb}{0.969, 0.969, 0.969}\color{fgcolor}\begin{figure}[!hbp]

{\centering \includegraphics[width=0.95\textwidth]{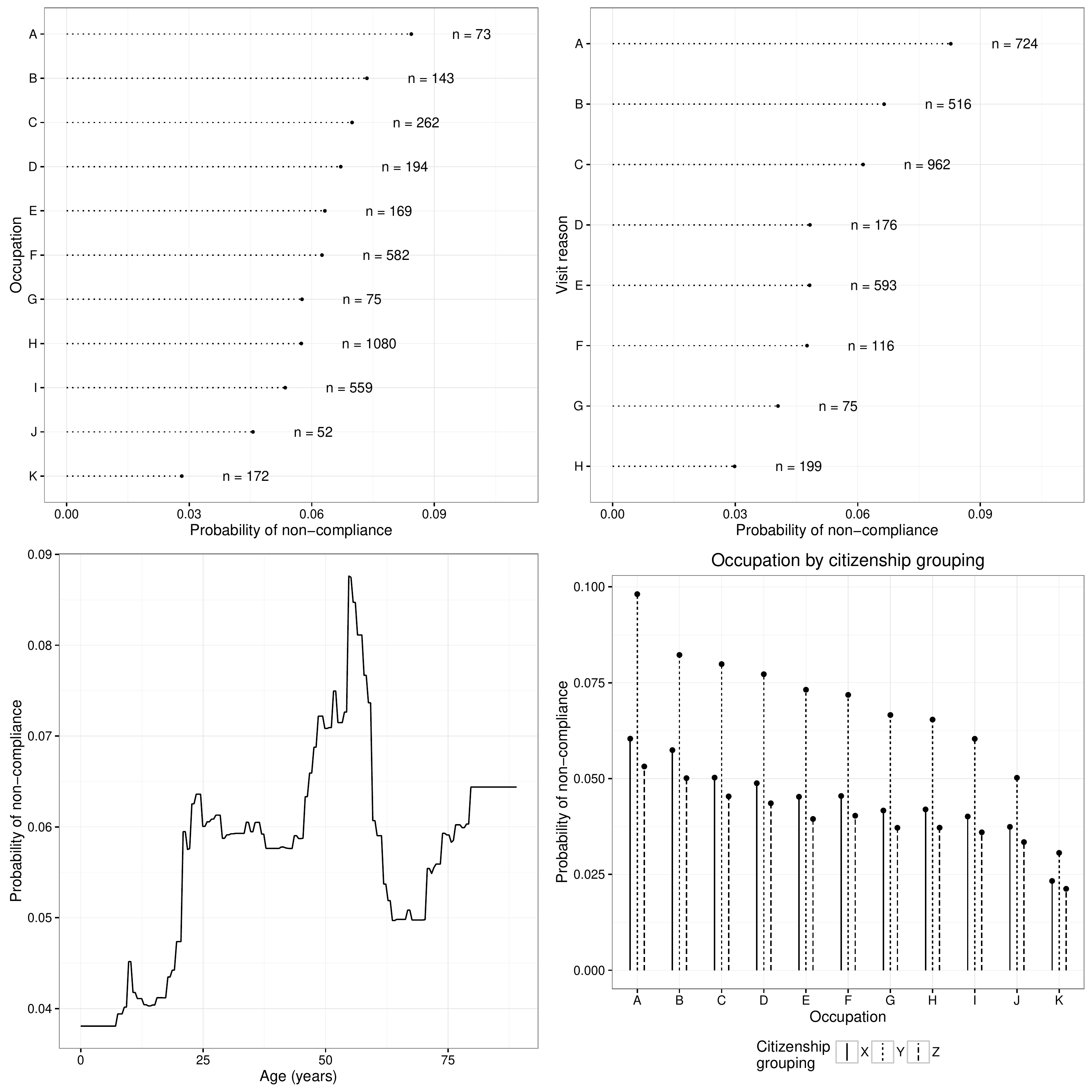} 

}

\caption[Marginal effect of selected passenger traits in predicting biosecurity risk material non-compliance in a GBM model fit to the full dataset]{Marginal effect of selected passenger traits in predicting biosecurity risk material non-compliance in a GBM model fit to the full dataset.}\label{fig:margEffect}
\end{figure}

\end{knitrout}

\section{Discussion}
\label{sec:discussion}

The results of the present case study show that for this dataset --- using only information collected from IPCs --- detection of non-compliance can be improved using statistical profiling. Importantly, we demonstrate this result holds relative to both random and manual selection. The predictive ability of most methods trialled was good, despite having a small dataset for modelling (both in terms of number of observations and variables). We conclude that data collected from IPCs is sufficient for constructing beneficial models that predict non-compliance for a wide range of model-fitting methods.

For any reasonable screening rate (up to 50\% screening of all passengers), the statistical profiling methods were more efficient than manual profiling (Figure~\ref{fig:modsHitRateInd}). This observation is important in the context of increasing passenger clearances as discussed in the introduction. With the increase in passenger clearances, either resources are spread more thinly on the ground, resulting in a lower screening rate, and/or budgets need to be increased for the extra personnel required to maintain the status quo. The results of the present case study indicate that it may in fact be possible to decrease the screening rate, whilst still maintaining a suitable level of biosecurity risk material interceptions. The end result of such a change, were it to be implemented, is likely to be a more efficient use of ground staff to meet the currently challenging environment.

In situations where automated screening --- that is, electronic screening prior to entry --- is not possible, it would be advantageous to have approximate profiling rules that could be used based upon a visual profile only. We fit the models in two stages against this contingency. The first stage used passenger traits amenable to manual decision making: age and sex, whilst the second stage added in variables that would only be available upon inspection of the IPCs. Models fit using all available data from the IPCs (Stage 2 models) clearly outperformed the Stage 1 models (Figure~\ref{fig:modsAUC}). We thus recommend that profiles be created using all available data, and be automated if possible.

As demonstrated in Figure~\ref{fig:modsHitRateInd} however, the Stage 1 models were still more efficient than current manual profiling, up to screening rates of approximately 30\% using the \textbf{GBM-caret} method. These results suggest that if automated processing is not possible in the near future, profiles can still be constructed from Stage 1 models --- for example by creating simple scoring rules based on effect sizes and variation, similar to the nomograms created in the medical literature for survival prediction (e.g., \cite{Callegaro2016-vp}). Such simple scoring rules still require time to process and if to be implemented via a visual screening approach, may be too cognitively demanding. However the benefit of doing so would be that direct officer engagement with passengers is lowered, as no interrogation of IPCs is required. Again, this has the flow-on effect of efficiency. Until such a time that information provided on IPCs is collected electronically, and likewise screening is automated, this approach should be trialled further.

This analysis enabled testing of a long-held informal belief among biosecurity inspectors that the declaration status of a passenger did not necessarily indicate their compliance status. That is, the declaration of one type of BRM does not reliably preclude the possibility that other BRM may be present.  If declarant passengers presented lower risk than non-declarant passengers then we would expect declaration status to play an important role in the model. Figure~\ref{fig:varImp} shows, however, that declaration status was not as important as age, occupation or visit reason in predicting non-compliance. We conclude that declarant passengers may still present risk of undetected BRM, with the caveat of course that the declaration of BRM still reduces the material risk.

The case study presented was restricted to only one dataset, collected during a fixed time period of two months, and was based upon a single flight number arriving at a single airport. Creating prediction rules to be used operationally was not the focus, and further research should investigate multiple flights covering different arrival times, and multiple entry points. Country/port of origin of the passenger, and time of departure/arrival could be investigated in expanded datasets. However, the clear performance advantages of statistical profiling compared with manual profiling as demonstrated in this study are only likely to improve with richer, expanded datasets.

Finally, in order to maintain current information about all cohorts of passengers, a proportion of the lower risk passengers should also be screened in addition to all the higher risk passengers identified by statistical profiling \citep{Robinson2011-aw}. This extra inspection is for three reasons: i) it allows the monitoring of changes in risk for the lower predicted risk groups; ii) any update to the models used for statistical profiling will require observations from the lower risk group in order to form unbiased predictions, and iii) it may act as a deterrent to non-targeted passengers.

In conclusion, we recommend that authorities investigate the use of targeted screening using statistical profiling for biosecurity risk material. However, larger, more comprehensive datasets should be assessed before implementation.

\printbibliography

\clearpage

\appendix

\section{Appendices}

\subsection{Model descriptions}
\label{sec:appendix-mods}

Table \ref{tab:modPreds} provides a full description of the models and the methods used for selecting tuning parameters (where required). All models were fit in \texttt{R} \citep{R_Core_Team2016-zb}, with package \texttt{caret} \citep{R-caret} used to select tuning parameters for the GBM \citep{R-gbm}, neural network \citep{R-nnet} and random forest \citep{R-randomForest} models. Tuning parameters for each prediction model were chosen by maximising predictive log-loss through repeated tenfold cross-validation, performed with five repetitions. Package \texttt{rstan} \citep{R-rstan} was used to fit the Bayesian regression models via the MCMC sampler \texttt{stan} \citep{Carpenter2016-oe}. The logistic regression model was fit using package \texttt{mgcv} \citep{Wood2006-rn}. As discussed in Section~\ref{prediction-models-compared}, two separate sub-models were fit, based on passenger traits that do not need the IPC, and those that do. Table~\ref{tab:modPreds} provides detail for the second stage of models; the first stage is identical, but with a limited set of passenger traits used for modelling.

\begin{longtable}{p{0.15\linewidth}p{0.80\linewidth}}
  \caption{Description of prediction models used for model comparison. If not specified, tuning parameters for each prediction model are chosen by maximising (predictive) log-loss through repeated tenfold cross-validation, performed with five repetitions. Appendix \ref{sec:appendix-A} provides a brief overview of the tuning parameters for each method.}\label{tab:modPreds} \\
  \toprule
  \textbf{GAM} & A logistic regression model. Occupation, visit reason and citizen group enter the model as categorical variables (see Table \ref{tab:Vars}), whilst passenger age is modelled using thin-plate splines, with smoothing parameter selected via generalised cross-validation as implemented in the \texttt{mgcv} package. No interactions are included, and no variable selection or other model tuning is performed. \\
  \textbf{RF-caret} & A random forest model with tuning parameter mtry -- the number of randomly sampled split candidates $\in$ (2, 12, 23). \\
  \textbf{GBM-custom} & A GBM with custom feature selection. Firstly, categorical variables are converted to dummy variables. Repeated subset sampling is then used to split the data into five training/testing datasets with a 70:30 split. For each (dummy) variable, a GBM is fit with tuning parameters set to: learning rate = 0.005; number of trees = 200; interaction depth = 2; and weights = 7 or 1 for observations that are non-compliant or compliant respectively. AUC is calculated for each individual model fit, and any variable with AUC \textless{} 0.51 is removed from further consideration. A GBM is fit to the remaining variables, and any variable with relative influence = 0 is removed. The \textbf{GBM-custom} model is a GBM model on the remaining variables. \\
  \textbf{GBM-caret} & A GBM model with tuning parameters optimised over a grid of: learning rate $\in$ (0.1, 0.01, 0.005); number of trees $\in$ (700, 850, 1000); and interaction depth $\in$ (1, 2, 3). \\
  \textbf{NN-caret} & A (single hidden layer) neural network model with tuning parameters over a grid of: size/number of units in hidden layer $\in$ (1, 3, 5); and weight decay $\in$ (0, 0.1, 0.0001). \\
  \textbf{Bayes-normal} & Bayesian (shrinkage) logistic regression with normal priors on the regression coefficients. We assume a binomial model: \\
  ~ & {$\!\begin{aligned}
      \Pr\big(Y_{i}=1\big|X_{i}\big) & = \text{logit}^{-1}\left(\mu + \beta_{m}M_{i} + \beta_{d}D_{i} + f(A_{i}) + \alpha_{o[i]} + \alpha_{v[i]} + \alpha_{c[i]}\right) \\
      f(A_{i}) & = \beta_{a}A_{i} + \sum_{k=1}^{K}\xi_{k}z_{ik} \\
      \mu & \sim N(0, 10) \\
      \beta_{j} & \sim N(0, \sigma_{s}^{2})\text{, for } j = m, d, a \\
      \xi_{k} & \sim N(0, \sigma_{a}^{2}) \\
      \alpha_{k[i]} & \sim N(0, \sigma_{s}^{2})\text{, for } k = o, v, c \\
      \sigma_{s} & \sim \text{half-}t_{1}(0, 1.0) \\
      \sigma_{a} & \sim \text{half-}t_{1}(0, 2.5)
    \end{aligned}$} \\
  ~ & where $m,d$ denote male passengers and passengers who declared they had BRM respectively, $a$ denotes passenger age, and $o,v,c$ denote the categorical variables of passenger occupation, reason for visiting and citizenship group respectively (see Table~\ref{tab:Vars}). $f(\cdot)$ is a thin-plate spline, where the parameters $\xi_{k}$ control the amount of shrinkage to the linear term \citep[construction of the $z_{ik}$ follows][]{Crainiceanu2005-uh}. Here the notation $\alpha_{k[i]}$ represents the coefficients for the \textit{varying levels} of the $k^{\text{th}}$ categorical variable \citep[following][]{Gelman2006-jx}. \\
  \textbf{Bayes-lasso} & Bayesian (shrinkage) logistic regression with Laplace priors \citep[Bayesian lasso model, e.g.][]{Park2008-la,Peltola2014-bp}. The binomial model is the same as for the \textbf{Bayes-normal} model, with the following changes to the coefficient priors: \\
  ~ & {$\!\begin{aligned}
      \beta_{j} & \sim N(0, \sigma_{s}^{2}\sigma_{j}^{2})\text{, for } j = m, d \\
      \alpha_{k[i]} & \sim N(0, \sigma_{s}^{2}\sigma_{k}^{2})\text{, for } k = o, v, c, a \\
      \sigma_{s} & \sim \text{half-}t_{1}(0, 1.0) \\
      \sigma_{j}^{2} & \sim \text{Exp}(1)\text{, for } j = m, d \\
      \sigma_{k}^{2} & \sim \text{Exp}(1)\text{, for } k = o, v, c, a
    \end{aligned}$} \\
  \bottomrule
\end{longtable}

\subsection{Tuning parameters}
\label{sec:appendix-A}

In this appendix we provide some detail on the tuning parameters that are available in the prediction methods of Section \ref{sec:methods}. \citet{Hastie2009-ns} provides an excellent overview of the methods and details of the tuning parameters.

\begin{longtable}{p{0.30\linewidth}p{0.65\linewidth}}
  \toprule
  \multicolumn{2}{l}{\textbf{Gradient boosting machine}} \\
  Number of trees & Controls the number of weak learners in the model. More trees lead to reduced error on the training set, but result in overfitting. \\
  Learning rate, $\nu \in (0, 1]$ & Provides regularisation by scaling the contribution of each weak learner to the current model fit. Lower values provide greater generalisation at a computational cost. \\
  Interaction depth & Controls the maximum degree of interactions. A value of 2, for example, allows two-way interactions. \\
  \multicolumn{2}{l}{\textbf{Neural network}} \\
  Weight decay & Controls the amount of shrinkage of the unknown weights towards 0. A weight decay value of 0 implies no shrinkage. \\
  Size/number of units in hidden layer & The number of \textit{derived features} in the hidden layer. These act similar to a basis expansion of the original predictors. \\
  \multicolumn{2}{l}{\textbf{Random forest}} \\
  Number of randomly sampled split candidates & Number of predictor variables that are randomly sampled at each split of the weak learner. \\
  \bottomrule
\end{longtable}

\end{document}